\documentclass[12pt]{article}
\usepackage{amsmath,amssymb,amsthm,amsxtra,overpic,bbm,bm,epsfig,ulem}
\usepackage{color,ulem}
\def\thefootnote{\fnsymbol{footnote}}

\begin{document}

\vspace{0.2cm}

\begin{center}
{\Large\bf The $\mu$-$\tau$ reflection symmetry of Dirac neutrinos \\ and
its breaking effect via quantum corrections}
\end{center}

\vspace{0.2cm}

\begin{center}
{\bf Zhi-zhong Xing$^{a, b}$},
~ {\bf Di Zhang$^{a}$\footnote{Email: dizhang@mails.ccnu.edu.cn},
~ {\bf Jing-yu Zhu}$^{a}$}\footnote{Email: zhujingyu@ihep.ac.cn} \\
{$^a$Institute of High Energy Physics, and School of Physical
Sciences, \\ University of Chinese Academy of Sciences, Beijing 100049, China \\
$^b$Center for High Energy Physics, Peking University, Beijing
100080, China}
\end{center}

\vspace{1.5cm}

\begin{abstract}
Given the Dirac neutrino mass term, we explore the constraint
conditions which allow the corresponding mass matrix
to be invariant under the $\mu$-$\tau$ reflection transformation,
leading us to the phenomenologically favored predictions
$\theta^{}_{23} = \pi/4$ and $\delta = 3\pi/2$ in the standard
parametrization of the $3\times 3$ lepton flavor mixing matrix.
If such a flavor symmetry is realized at a superhigh energy scale
$\Lambda^{}_{\mu\tau}$, we investigate how it is
spontaneously broken via the one-loop renormalization-group equations
(RGEs) running from $\Lambda^{}_{\mu\tau}$ down to the Fermi scale
$\Lambda^{}_{\rm F}$. Such quantum corrections to the neutrino masses
and flavor mixing parameters are derived, and an analytical
link is established between the Jarlskog invariants of CP
violation at $\Lambda^{}_{\mu\tau}$ and $\Lambda^{}_{\rm F}$.
Some numerical examples are also presented in both the minimal
supersymmetric standard model and the type-II two-Higgs-doublet model, to
illustrate how the octant of $\theta^{}_{23}$, the quadrant of
$\delta$ and the neutrino mass ordering are correlated with one another
as a result of the RGE-induced $\mu$-$\tau$ reflection symmetry breaking
effects.
\end{abstract}

\begin{flushleft}
\hspace{0.8cm} PACS number(s): 14.60.Pq, 11.30.Hv, 11.10.Hi.
\end{flushleft}

\def\thefootnote{\arabic{footnote}}
\setcounter{footnote}{0}

\newpage

\section{Introduction}

The discoveries of solar, atmospheric, reactor and accelerator
neutrino oscillations \cite{PDG} have demonstrated that the standard model
(SM) of electroweak interactions is incomplete and must be extended
in a proper way so as to accommodate
tiny neutrino masses and significant lepton flavor mixing. The
simplest way to do so is to introduce three right-handed
(or $\rm SU(2)$-singlet) neutrino
fields $N^{}_{\alpha \rm R}$ (for $\alpha = e, \mu, \tau$) into
the SM and write out a gauge-invariant, Lorentz-invariant and
lepton-number-conserving mass term of the form
\begin{eqnarray}
-\mathcal{L}^{}_{\rm Dirac} = \overline{\ell^{}_{\rm L}}
Y^{}_{\nu} \tilde{H} N^{}_{\rm R} + {\rm h.c.} \; ,
%  (1)
\end{eqnarray}
where $\tilde{H}={\rm i}\sigma^{}_{2} H^{\ast}$ with $H$ being
the SM Higgs doublet, $\ell^{}_{\rm L}$ denotes the left-handed
lepton doublet column vector, and $N^{}_{\rm R}$ represents the right-handed
neutrino column vector with the $N^{}_{\alpha \rm R}$ components.
After spontaneous gauge symmetry breaking, the above Dirac neutrino
mass term turns out to be
%%%%%%%%%%%%%%%%%%%%%%%%%%%%%%%%%%%%%%%%%%%%%%%%%%%%%%%%%%%%%%%%%%%%%%
\footnote{If the minimal supersymmetric standard model (MSSM) 
is concerned, the charged-lepton
and neutrino sectors are associated with the Higgs doublets
$H^{}_1$ (with the hypercharge $+1/2$ and the vacuum expectation value
$v\cos\beta/\sqrt{2}$) and $H^{}_2$ (with the hypercharge $-1/2$ and
the vacuum expectation value $v \sin\beta/\sqrt{2}$), respectively
\cite{Mei}. But for the type-II two-Higgs-doublet model (2HDM), the Higgs
doublet $H^{}_1$ is coupled to both the charged-lepton
and neutrino sectors \cite{Branco}. These two interesting scenarios will be
used to illustrate quantum corrections to the $\mu$-$\tau$ reflection
symmetry in section 4.}
%%%%%%%%%%%%%%%%%%%%%%%%%%%%%%%%%%%%%%%%%%%%%%%%%%%%%%%%%%%%%%%%%%%%%%%
\begin{eqnarray}
-\mathcal{L}^{\prime}_{\rm Dirac} = \overline{\nu^{}_{\rm L}}
M^{}_\nu N^{}_{\rm R} + {\rm h.c.} \; ,
%  (2)
\end{eqnarray}
where $M^{}_\nu = Y^{}_{\nu}\langle H\rangle$ with
$\langle H\rangle = v/\sqrt{2}$ and $v \simeq 246$ GeV. The three neutrino
masses $m^{}_i$ (for $i=1,2,3$) can therefore be achieved from
diagonalizing $M^{}_\nu$ if its texture is specified in a given model,
but the smallness of $m^{}_i$ is not really explained in this manner.
While many theorists believe that the neutrinos should be Majorana
fermions \cite{M}, by which their small masses can be naturally
understood via a seesaw mechanism \cite{S1,Xing09},
the simplicity of the Dirac neutrino mass generation
mechanism {\it do} attract quite a lot of attention \cite{D1,D2}.
Before the Majorana nature of massive neutrinos is ultimately determined
with the help of a measurement of the neutrinoless double-beta decay
or other lepton-number-violating processes \cite{Giunti},
it makes sense to study the phenomenology of Dirac neutrinos as well.

Assuming the massive neutrinos to be the Dirac fermions, we shall begin
with Eq. (2) to explore the $\mu$-$\tau$ reflection symmetry of
$\mathcal{L}^{\prime}_{\rm Dirac}$ and the resulting texture of $M^{}_\nu$
in the basis where the flavor eigenstates of three charged leptons are
identified with their mass eigenstates.
The motivation for this study is simply because such a flavor symmetry
may naturally lead us to the phenomenologically favored predictions
$\theta^{}_{23} = \pi/4$ and $\delta =3\pi/2$ in the standard
parametrization of the $3\times 3$ Pontecorvo-Maki-Nakagawa-Sakata (PMNS)
lepton flavor mixing matrix $U$ \cite{PMNS} that is used to diagonalize
$M^{}_\nu M^\dagger_\nu$. Provided the $\mu$-$\tau$ reflection symmetry
is realized at a superhigh energy scale $\Lambda^{}_{\mu\tau}$, we shall
investigate how it is spontaneously broken due to the running of
$M^{}_\nu$ from $\Lambda^{}_{\mu\tau}$ down to the Fermi scale
$\Lambda^{}_{\rm F} \sim v \sim 10^2$ GeV through the one-loop
renormalization-group equations (RGEs) in the framework of either the
MSSM or the type-II 2HDM. Such quantum corrections to the
three neutrino masses and four flavor mixing parameters will be derived,
and an analytical link will be established between the Jarlskog
invariants of leptonic CP violation at $\Lambda^{}_{\mu\tau}$ and
$\Lambda^{}_{\rm F}$. We shall also present some numerical examples
in both the MSSM and the type-II 2HDM to illustrate how the octant of
$\theta^{}_{23}$, the quadrant of $\delta$ and the neutrino mass
ordering are correlated with one another as a result of the
RGE-triggered $\mu$-$\tau$ reflection symmetry breaking effects.

The content of this work is new in several aspects.
First, applying the $\mu$-$\tau$
reflection symmetry to the Dirac neutrino mass term, in which $M^{}_\nu$
is in general neither symmetric nor Hermitian, has not been
tried before. Second, the integral form of the RGE corrections to
$M^{}_\nu$ is derived for the first time, so is the integral form
of the RGE effects on the neutrino masses and flavor mixing
parameters. Third, a concise analytical relationship between the Jarlskog
invariants of CP violation at $\Lambda^{}_{\mu\tau}$ and $\Lambda^{}_{\rm F}$
is derived for the first time. Fourth, a comparison is made between the
MSSM and the type-II 2HDM, which leads to the opposite deviations of
$\theta^{}_{12}$, $\theta^{}_{13}$, $\theta^{}_{23}$ and $\delta$
from their corresponding values in the $\mu$-$\tau$ reflection
symmetry limit.

The remaining parts of this paper are organized as follows. In
section 2 we shall find out the constraint conditions
which allow the Dirac neutrino mass matrix $M^{}_\nu$
to be invariant under the $\mu$-$\tau$ reflection transformation.
Section 3 is devoted to the derivation of the integral form of
the RGE corrections to $M^{}_\nu M^\dagger_\nu$ when it runs from
$\Lambda^{}_{\mu\tau}$ down to $\Lambda^{}_{\rm F}$, and to the
derivation of an analytical relationship between the Jarlskog invariants at
$\Lambda^{}_{\mu\tau}$ and $\Lambda^{}_{\rm F}$. In section 4
we calculate the RGE-induced corrections to the neutrino masses and
flavor mixing parameters in a perturbation way, and illustrate
their salient features by taking a few numerical examples in
both the MSSM and the type-II 2HDM. Finally,
we summarize our main results and make a conclusion in section 5.

\section{$\mu$-$\tau$ reflection symmetry}

Given the Dirac neutrino mass term in Eq. (2), let us consider
the following transformations of the six neutrino fields
%%%%%%%%%%%%%%%%%%%%%%%%%%%%%%%%%%%%%%%%%%%%%%%%%%%%%%%%%%%%%%
\footnote{In this work we focus on a possible $\mu$-$\tau$ reflection
symmetry of the Dirac neutrino mass matrix {\it after} spontaneous
gauge symmetry breaking. Otherwise, the neutrino field transformations
made in Eq. (3) would affect some other parts of the Lagrangian of the
electroweak interactions. To build a consistent lepton mass model
with the $\mu$-$\tau$ flavor symmetry in the neutrino sector
instead of the charged-lepton sector, one should introduce some
extra scalar fields coupling to the two sectors in a different way
\cite{Mohapatra}. But here we simply assume that the $\mu$-$\tau$
reflection symmetry does not apply to the charged-lepton sector.
In this sense the invariance of ${\cal L}^\prime_{\rm Dirac}$
under the transformations in Eq. (3) can just serve as
a phenomenological guiding principle to obtain the special texture
of $M^{}_\nu$ in Eq. (9).}:
%%%%%%%%%%%%%%%%%%%%%%%%%%%%%%%%%%%%%%%%%%%%%%%%%%%%%%%%%%%%%%
\begin{eqnarray}
&& \nu^{}_{e \rm L} \leftrightarrow \nu^{\rm c}_{e \rm L} \; , \hspace{0.1cm}
\quad \quad N^{}_{e \rm R} \leftrightarrow N^{\rm c}_{e \rm R} \; ,
\nonumber \\
&& \nu^{}_{\mu \rm L} \leftrightarrow \nu^{\rm c}_{\tau \rm L} \; ,
\quad \quad N^{}_{\mu \rm R} \leftrightarrow N^{\rm c}_{\tau \rm R} \; ,
\nonumber \\
&& \nu^{}_{\tau \rm L} \leftrightarrow \nu^{\rm c}_{\mu \rm L} \; ,
\quad \quad N^{}_{\tau \rm R} \leftrightarrow N^{\rm c}_{\mu \rm R} \; ,
%   (3)
\end{eqnarray}
where $\nu^{\rm c}_{\alpha \rm L} \equiv {\cal C}
\overline{\nu^{}_{\alpha \rm L}}^{\rm T}$ and
$N^{\rm c}_{\alpha \rm L} \equiv {\cal C}
\overline{N^{}_{\alpha \rm L}}^{\rm T}$ (for $\alpha = e, \mu, \tau$)
with T denoting the transpose and
$\cal C$ being the charge-conjugation operator and
satisfying ${\cal C}^{-1} = {\cal C}^\dagger = {\cal C}^{\rm T} =
-{\cal C}$ \cite{XZ2011}. Under such transformations, Eq. (2) turns out to be
\begin{eqnarray}
-\mathcal{L}^{'}_{\rm Dirac} \hspace{-0.2cm} & = & \hspace{-0.2cm}
\overline{\nu^{\rm c}_{\rm L}} S
M^{}_\nu S N^{\rm c}_{\rm R} + \overline{N^{\rm c}_{\rm R}} S
M^{\dagger}_\nu S \nu^{\rm c}_{\rm L}
\nonumber \\
\hspace{-0.2cm} & = & \hspace{-0.2cm}
-\nu^{\rm T}_{\rm L} S M^{}_\nu S \overline{N^{}_{\rm R}}^{\rm T}
- N^{\rm T}_{\rm R} S M^{\dagger}_\nu S \overline{\nu^{}_{\rm L}}^{\rm T}
\nonumber \\
\hspace{-0.2cm} & = & \hspace{-0.2cm}
\overline{\nu^{}_{\rm L}} S M^{\ast}_\nu S N^{}_{\rm R}
+ \overline{N^{}_{\rm R}} S M^{\rm T}_\nu S \nu^{}_{\rm L} \; ,
%   (4)
\end{eqnarray}
in which the property of ${\cal L}^\prime$ as a Lorentz scalar has been
used, and
\begin{eqnarray}
S = \left( \begin{matrix} 1 & 0 & 0 \cr 0 & 0 & 1 \cr 0 & 1 & 0
\end{matrix} \right) \; .
%   (5)
\end{eqnarray}
If ${\cal L}^\prime$ is required to be invariant under the above
$\mu$-$\tau$ reflection transformations \cite{HS}, then the Dirac
neutrino mass matrix
\begin{eqnarray}
M^{}_\nu \equiv \left( \begin{matrix} \langle m\rangle^{}_{ee} &
\langle m\rangle^{}_{e \mu} & \langle m\rangle^{}_{e \tau} \cr
\langle m\rangle^{}_{\mu e} &
\langle m\rangle^{}_{\mu \mu} & \langle m\rangle^{}_{\mu \tau} \cr
\langle m\rangle^{}_{\tau e} &
\langle m\rangle^{}_{\tau \mu} & \langle m\rangle^{}_{\tau \tau}
\end{matrix} \right) \; .
%   (6)
\end{eqnarray}
must satisfy the relationship
\begin{eqnarray}
M^{}_\nu = S M^{\ast}_\nu S \; .
%   (7)
\end{eqnarray}
In other words, the elements of $M^{}_\nu$ must satisfy
\begin{eqnarray}
&& \langle m\rangle^{}_{ee} = \langle m\rangle^{\ast}_{ee} \; ,
\quad \quad \langle m\rangle^{}_{e\mu} = \langle m\rangle^{\ast}_{e\tau} \; ,
\nonumber \\
&& \langle m\rangle^{}_{\mu e} = \langle m\rangle^{\ast}_{\tau e} \; ,
\quad \quad \hspace{-0.05cm}
\langle m\rangle^{}_{\mu\tau} = \langle m\rangle^{\ast}_{\tau\mu} \; ,
\hspace{1cm}
\nonumber \\
&& \langle m\rangle^{}_{\mu\mu} = \langle m\rangle^{\ast}_{\tau\tau} \; .
%   (8)
\end{eqnarray}
Then the texture of $M^{}_\nu$ can be simply parametrized as
\begin{eqnarray}
M^{}_\nu = \left( \begin{matrix} a & b & b^{\ast} \cr
e & c & d \cr e^{\ast} & d^{\ast} & c^{\ast}
\end{matrix} \right) \; ,
%   (9)
\end{eqnarray}
where $a$ is real, and the other four parameters are in general complex.
To diagonalize $M^{}_\nu$ in Eq. (9), one may do
a bi-unitary transformation of the form
\begin{eqnarray}
U^\dagger M^{}_\nu Q = \hat{M}^{}_\nu \; ,
%     (10)
\end{eqnarray}
where $U$ and $Q$ are the unitary matrices, and
$\hat{M}^{}_\nu \equiv {\rm Diag}\{m^{}_1, m^{}_2, m^{}_3\}$
with $m^{}_i$ (for $i=1,2,3$) being the neutrino masses. In the
basis where the flavor eigenstates of three charged
leptons are identified with their mass eigenstates, the unitary
matrix $U$ is just the PMNS flavor mixing matrix which manifests
itself in the leptonic weak charged-current interactions.

It proves more convenient to consider the Hermitian matrix
\begin{eqnarray}
H^{}_\nu \hspace{-0.2cm} & \equiv & \hspace{-0.2cm}
M^{}_\nu M^{\dagger}_\nu = U \hat{M}^{2}_\nu U^\dagger \;
\nonumber \\
\hspace{-0.2cm} & = & \hspace{-0.2cm}
\left( \begin{matrix} A & B & B^{\ast} \cr B^{\ast}
& C & D \cr B & D^{\ast} & C \end{matrix} \right) \; ,
\hspace{1cm}
%     (11)
\end{eqnarray}
where
\begin{eqnarray}
A \hspace{-0.2cm} & = & \hspace{-0.2cm} a^{2} + 2|b|^{2} \; ,
\nonumber \\
B \hspace{-0.2cm} & = & \hspace{-0.2cm} a e^{\ast} + b c^{\ast}
+ b^{\ast} d^{\ast} \; ,
\nonumber \\
C \hspace{-0.2cm} & = & \hspace{-0.2cm} |e|^{2} + |c|^{2} + |d|^{2} \; ,
\hspace{1cm}
\nonumber \\
D \hspace{-0.2cm} & = & \hspace{-0.2cm} e^{2} + 2 c d \; .
%  (12)
\end{eqnarray}
Moreover, let us parametrize $U$ as $U \equiv P V$, where
$P = {\rm Diag}\{e^{{\rm i} \phi^{}_e}, e^{{\rm i} \phi^{}_\mu},
e^{{\rm i} \phi^{}_\tau}\}$ is an unphysical phase matrix
associated with the charged-lepton fields
%%%%%%%%%%%%%%%%%%%%%%%%%%%%%%%%%%%%%%%%%%
\footnote{Note that these unphysical phases should not be ignored
{\it in the course} of deriving the RGEs of the neutrino masses and flavor mixing
parameters, as one can see in section 4. When using $U=PV$ to reconstruct
the texture of $H^{}_\nu$, we find that
$\phi^{}_\mu + \phi^{}_\tau = 2 \phi^{}_e$ must be satisfied, as required by the
$\mu$-$\tau$ reflection symmetry.},
%%%%%%%%%%%%%%%%%%%%%%%%%%%%%%%%%%%%%%%%%%
and
\begin{eqnarray}
V = \left(
\begin{matrix}
c^{}_{12} c^{}_{13} & s^{}_{12} c^{}_{13} &
s^{}_{13} e^{-{\rm i} \delta} \cr -s^{}_{12} c^{}_{23} - c^{}_{12}
s^{}_{13} s^{}_{23} e^{{\rm i} \delta} & c^{}_{12} c^{}_{23} -
s^{}_{12} s^{}_{13} s^{}_{23} e^{{\rm i} \delta} & c^{}_{13}
s^{}_{23} \cr -s^{}_{12} s^{}_{23} + c^{}_{12} s^{}_{13} c^{}_{23}
e^{{\rm i} \delta} & c^{}_{12} s^{}_{23} + s^{}_{12} s^{}_{13}
c^{}_{23} e^{{\rm i} \delta} & - c^{}_{13} c^{}_{23} \cr
\end{matrix} \right) \;
%     (13)
\end{eqnarray}
with $c^{}_{ij} \equiv \cos\theta^{}_{ij}$ and
$s^{}_{ij} \equiv \sin\theta^{}_{ij}$ (for $ij=12,13,23$). At a given
energy scale, one may rotate away $P$ and then express the four flavor
mixing parameters of $V$ in terms of the elements of 
\begin{eqnarray}
\overline{H}^{}_\nu \hspace{-0.2cm} & \equiv & \hspace{-0.2cm}
P^\dagger H^{}_\nu P = V \hat{M}^{2}_\nu V^\dagger \;
\nonumber \\
\hspace{-0.2cm} & = & \hspace{-0.2cm}
\left( \begin{matrix} A & \overline{B} & \overline{B}^{\ast} \cr \overline{B}^{\ast}
& C & \overline{D} \cr \overline{B} & \overline{D}^{\ast} & C \end{matrix} \right) \; ,
\hspace{1cm}
%     (14)
\end{eqnarray}
where $\overline{B} = B e^{{\rm i} (\phi^{}_{\mu} - \phi^{}_e)}$ and
$\overline{D} = D e^{{\rm i} (\phi^{}_{\tau} - \phi^{}_\mu)}$. In
this way the unphysical phases hidden in $B$ and $D$ will be 
cancelled by 
$\phi^{}_\mu - \phi^{}_e$ and $\phi^{}_\tau - \phi^{}_\mu$, 
respectively. Then we do a similar diagonalization of 
$\overline H^{}_{\nu}$ as that done in Ref. \cite{Yasue} and obtain
\begin{eqnarray}
&& \theta^{}_{12} = \frac{1}{2} \arctan \left[ 2 \frac{\left|{\rm Re}
\overline{B} \right| \sqrt{2 \left( {\rm Re} \overline{B} \right)^2 +
\left( {\rm Im} \overline{D}\right)^2 }}
{\left| {\rm Im} \overline{B} \ {\rm Im} \overline{D}
- 2 {\rm Re} \overline{B} \ {\rm Re} \overline{D} \right|} \right] \; ,
\hspace{1cm} \nonumber \\
&& \theta^{}_{13} = \arctan \left[
\frac{1}{\sqrt{2}} \left|\frac{{\rm Im} \overline{D}}
{{\rm Re} \overline{B}} \right| \right] \; ;
%     (15)
\end{eqnarray}
together with the typical predictions
\begin{eqnarray}
\theta^{}_{23} = \frac{\pi}{4} \; , \quad \quad
\delta = \frac{\pi}{2} \quad {\rm or} \quad \frac{3\pi}{2} \; .
%     (16)
\end{eqnarray}
These two numerical predictions, which have been well known for
the Majorana neutrino mass matrix with the $\mu$-$\tau$ reflection
symmetry \cite{XZ2016}, are now achieved in the Dirac case with
the same flavor symmetry. It is easy to see that Eq. (16) leads us to
the equalities
\begin{eqnarray}
\left|V^{}_{\mu 1}\right| = \left|V^{}_{\tau 1}\right| \; , \quad\quad
\left|V^{}_{\mu 2}\right| = \left|V^{}_{\tau 2}\right| \; , \quad\quad
\left|V^{}_{\mu 3}\right| = \left|V^{}_{\tau 3}\right| \; ,
%     (16)
\end{eqnarray}
which are sometimes referred to as the $\mu$-$\tau$ reflection
symmetry at the PMNS matrix level. One may therefore define the
asymmetries ${\cal A}^{}_i \equiv |V^{}_{\mu i}|^2 - |V^{}_{\tau i}|^2$
(for $i=1,2,3$) to measure the effects of $\mu$-$\tau$ symmetry
breaking in a rephasing-invariant way \cite{Luo}.

Of course, it is more fundamental to understand how the $\mu$-$\tau$
reflection symmetry of $M^{}_\nu$ or $H^{}_\nu$
can be spontaneously or explicitly
broken, both for the model-building purpose and for explaining
currently available neutrino oscillation data \cite{Zhu}. Following
the discussions about the $\mu$-$\tau$ symmetry breaking of the
Majorana neutrino mass matrix \cite{XZ2016,Zhao}, one can similarly
introduce the most general perturbation to the Dirac neutrino mass
matrix with the $\mu$-$\tau$ reflection symmetry.
But we find that it is more convenient to focus on the perturbation
to $H^{}_\nu$ in Eq. (11) instead of $M^{}_\nu$ in Eq. (9), simply
because the former is always Hermitian. In this case the perturbation
matrix $\Delta H^{}_\nu$ can also be arranged to be Hermitian,
and it can be decomposed into two parts: one part conserves
the original $\mu$-$\tau$ reflection symmetry and the other part
violates this symmetry. Namely,
\begin{eqnarray}
\Delta H^{}_\nu = \left(\begin{matrix} \delta^{}_{ee} & \delta^{}_{e\mu}
& \delta^{}_{e\tau} \cr \delta^{\ast}_{e\mu} & \delta^{}_{\mu\mu}
& \delta^{}_{\mu\tau} \cr \delta^{\ast}_{e\tau} & \delta^{\ast}_{\mu\tau}
& \delta^{}_{\tau\tau} \end{matrix}\right)
\hspace{-0.2cm} & = & \hspace{-0.2cm}
\frac{1}{2} \left(\begin{matrix} 2\delta^{}_{ee} & \delta^{}_{e\mu}
+ \delta^{\ast}_{e\tau} & \delta^{\ast}_{e\mu} + \delta^{}_{e\tau} \cr
\delta^{\ast}_{e\mu} + \delta^{}_{e\tau} & \delta^{}_{\mu\mu} +
\delta^{}_{\tau\tau} & 2\delta^{}_{\mu\tau} \cr \delta^{}_{e\mu} +
\delta^{\ast}_{e\tau} & 2\delta^{\ast}_{\mu\tau} & \delta^{}_{\mu\mu}
+\delta^{}_{\tau\tau} \end{matrix}\right)
\nonumber \\
\hspace{-0.2cm} & + & \hspace{-0.2cm}
\frac{1}{2} \left(\begin{matrix} {\bf 0} & \delta^{}_{e\mu} -
\delta^{\ast}_{e\tau} & \delta^{}_{e\tau} - \delta^{\ast}_{e\mu} \cr
\delta^{\ast}_{e\mu} - \delta^{}_{e\tau} & \delta^{}_{\mu\mu} -
\delta^{}_{\tau\tau} & {\bf 0} \cr \delta^{\ast}_{e\tau} - \delta^{}_{e\mu}
& {\bf 0} & \delta^{}_{\tau\tau} - \delta^{}_{\mu\mu} \end{matrix}\right) \; ,
%  (18)
\end{eqnarray}
where $\delta^{}_{ee}$, $\delta^{}_{\mu\mu}$ and $\delta^{}_{\tau\tau}$ are real,
and all the parameters are expected to be reasonably small in magnitude.
Because the symmetry-conserving part can be absorbed into $H^{}_\nu$
via a redefinition of its initial matrix elements, we are then left with
\begin{eqnarray}
H^{\prime}_\nu = H^{}_\nu + \Delta H^{}_\nu = \left(\begin{matrix} A^{\prime} & B^{\prime} \left(1+\epsilon^{}_{1}\right) & B^{\prime\ast}\left(1-\epsilon^{\ast}_{1}\right) \cr B^{\prime\ast}\left(1+\epsilon^{\ast}_{1}\right) & C^{\prime}\left(1+\epsilon^{}_{2}\right)
& D^{\prime} \cr B^{\prime}\left(1-\epsilon^{}_{1}\right) & D^{\prime\ast} & C^{\prime}\left(1-\epsilon^{}_{2}\right)  \end{matrix}\right) \; ,
%  (19)
\end{eqnarray}
where
\begin{eqnarray}
&& A^{\prime} = A + \delta^{}_{ee} \; , \quad \quad
B^{\prime} = B + \frac{\delta^{}_{e\mu} + \delta^{\ast}_{e\tau}}{2} \; ,
\nonumber \\
&& C^{\prime} = C + \frac{\delta^{}_{\mu\mu} + \delta^{}_{\tau\tau}}{2} \; ,
\quad \quad D^{\prime} = D + \delta^{}_{\mu\tau} \; \hspace{1.3cm}
%  (20)
\end{eqnarray}
and
\begin{eqnarray}
\epsilon^{}_{1} = \frac{\delta^{}_{e\mu} -
\delta^{\ast}_{e\tau}}{2B^{\prime}} \; ,
\quad \quad \quad \quad \hspace{-0.3cm}
\epsilon^{}_{2} = \frac{\delta^{}_{\mu\mu} -
\delta^{}_{\tau\tau}}{2C^{\prime}_{}} \; .
%  (21)
\end{eqnarray}
It is obvious that $\epsilon^{}_{1}$ and $\epsilon^{}_{2}$ are complex and
real, respectively. These two dimensionless parameters will vanish, if
$\Delta H^{}_\nu$ respects the $\mu$-$\tau$ reflection symmetry.

Although the above formulism can provide us with a generic picture of the
$\mu$-$\tau$ symmetry breaking, it has to be specified so as to see the
explicit symmetry-breaking effects. In the following we shall assume that
the $\mu$-$\tau$ reflection symmetry is realized at a superhigh energy
scale $\Lambda^{}_{\mu\tau}$, and examine its breaking at the Fermi
scale $\Lambda^{}_{\rm F}$ via the one-loop RGEs.

\section{RGE corrections to $H^{}_\nu$}

From the point of view of model building, a specific flavor symmetry is
usually realized at a superhigh energy scale where some fundamental new
physics beyond the SM can naturally manifest itself.
In this case the phenomenological consequences of such a flavor symmetry
should be confronted with the low-energy experimental data by running
the relevant physical quantities down to the Fermi scale
$\Lambda^{}_{\rm F}$ via the RGEs. In Ref. \cite{Luo} the one-loop
RGEs of the $\mu$-$\tau$ asymmetries ${\cal A}^{}_i$ of the PMNS matrix $U$
have been derived. Here we are going to derive the integral form of
the RGE corrections to $M^{}_\nu$ and $H^{}_\nu$.

The differential form of the one-loop RGE for the Dirac neutrino mass
matrix $M^{}_\nu$ in the framework of the MSSM or the 2HDM
is known as \cite{RGE,Chiang}
\begin{eqnarray}
16\pi^{2} \frac{{\rm d} M^{}_\nu}{{\rm d} t} = \left[ G
+ C_{\nu}^{} Y^{}_\nu Y^\dagger_\nu + C_{l}^{} Y^{}_{l}Y^{\dagger}_{l}
\right] M^{}_\nu \; ,
%  (22)
\end{eqnarray}
where $t \equiv \ln \left(\Lambda/\Lambda^{}_{\mu\tau}\right)$ with
$\Lambda$ being a renormalization scale,
$Y^{}_\nu$ and $Y^{}_l$ are the Yukawa coupling
matrices of the neutrinos and charged leptons, respectively. 
Given the MSSM, one has $C_{\nu}^{} =3$, $C_{l}^{} =1$, and 
$G \simeq -0.6 g^{2}_{1} - 3g^{2}_{2}+3y^{2}_{t}$
with $g^{}_{1,2}$ being the gauge couplings and $y^{}_t$ being
the top-quark Yukawa coupling in the $y^2_u \ll y^2_c \ll y^2_t$
approximation. If the type-II 2HDM is taken into account, one has
$C_{\nu}^{} =3/2$, $C_{l}^{} =-3/2$, and
$G \simeq -0.45 g^{2}_{1} - 2.25 g^{2}_{2} + 
y^2_{\tau} + 3y^{2}_{b}$ with $y_{\tau}^{}$ and
$y_{b}^{}$ being the tau-lepton and bottom-quark Yukawa couplings
in the $y^2_{e} \ll y^2_{\mu} \ll y^2_{\tau}$ and
$y^2_{d} \ll y^2_{s} \ll y^2_{b}$ approximations.
Since the neutrino masses $m^{}_i$ are extremely small as compared
with their charged partners, it is very safe to neglect the
$Y^{}_\nu Y^\dagger_\nu$ term in Eq. (22). In the basis that we
have chosen (i.e., the mass eigenstates of three charged leptons
are identified with their flavor eigenstates),
$Y^{}_l Y^\dagger_l = D^2_l \equiv {\rm Diag}\{y^2_e, y^2_\mu, y^2_\tau\}$
holds, where $y^2_\alpha = 2 \left(1 + \tan^2\beta\right) m^2_\alpha/v^2$
(for $\alpha =e, \mu, \tau$) with $\tan\beta$ being the ratio of the
vacuum expectation value of $H^{}_2$ to that of $H^{}_1$ in the MSSM or
the type-II 2HDM. Then Eq. (22) leads us to the RGE of $H^{}_\nu$ as follows:
\begin{eqnarray}
16\pi^{2}_{}\frac{{\rm d} H^{}_\nu}{{\rm d} t} = 2 \hspace{0.06cm}
G H^{}_\nu + D^2_l H^{}_\nu + H^{}_\nu D^2_l \; .
%  (23)
\end{eqnarray}
Integrating Eq. (23) from $\Lambda^{}_{\mu\tau}$ to $\Lambda^{}_{\rm F}$,
we immediately arrive at
\begin{eqnarray}
H^{\prime}_\nu = I^{2}_G T^{}_l H^{}_\nu T^{}_l \; ,
%  (24)
\end{eqnarray}
where $H^{}_\nu$ and $H^\prime_\nu$ are associated respectively with
the scales $\Lambda^{}_{\mu\tau}$ and $\Lambda^{}_{\rm F}$,
$T^{}_l \equiv {\rm Diag}\{I^{}_e, I^{}_\mu, I^{}_\tau\}$, and
the evolution functions are
\begin{eqnarray}
&& I^{}_G = \exp\left[ \frac{1}{16\pi^{2}}
\int^{t^\prime}_{0} G \ {\rm d} t \right] \; ,
\nonumber \\
&& I^{}_\alpha = \exp\left[ \frac{C^{}_l}{16\pi^{2}}
\int^{t^\prime}_{0} y^{2}_\alpha \ {\rm d} t \right] \; , \hspace{1.5cm}
%     (25)
\end{eqnarray} 
where $t^\prime \equiv \ln(\Lambda^{}_{\rm F}/\Lambda^{}_{\mu\tau})$,
and $\alpha$ runs over $e$, $\mu$ and $\tau$.
If one is more interested in the relationship
between $M^{\prime}_\nu$ at $\Lambda^{}_{\rm F}$ and $M^{}_\nu$ at
$\Lambda^{}_{\mu\tau}$, then it is straightforward to obtain
\begin{eqnarray}
M^{\prime}_\nu = I^{}_G T^{}_l M^{}_\nu \; ,
%  (26)
\end{eqnarray}
either from integrating Eq. (22) or from decomposing Eq. (24).
%%%%%%%%%%%%%%%%%%%%%%%%%%%%%%%%%%%%%%%%%%%%%%%%%%%%%%%%%%%%%%
%%%%%%%%%%%%%%%%%%%%% Figure 1 %%%%%%%%%%%%%%%%%%%%%%%%%%%%%%%
\begin{figure}[t]
	\includegraphics[width=16.5cm]{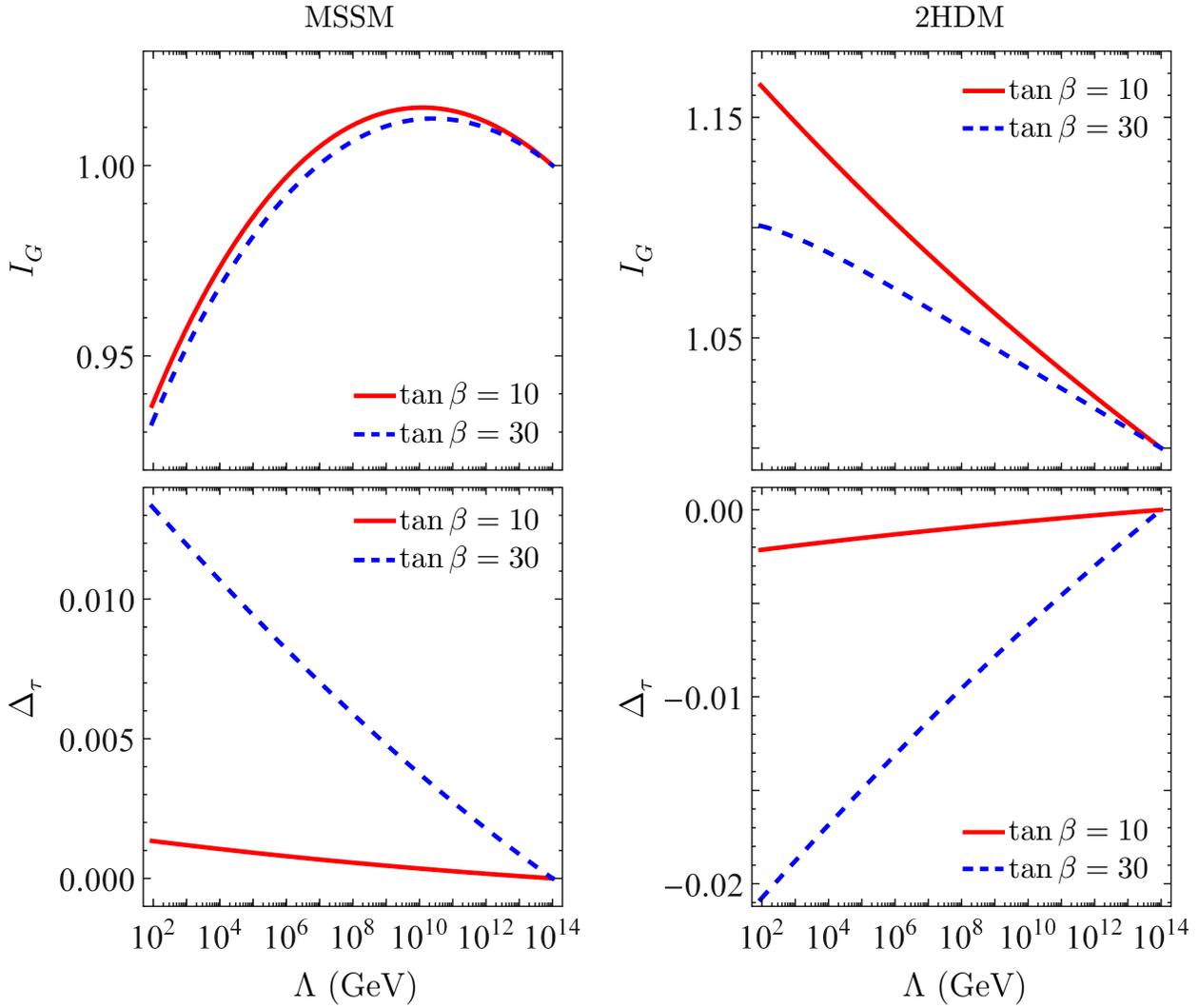}
	\caption{Changes of $I^{}_{G}$ and $\Delta^{}_{\tau}$ versus
the energy scale $\Lambda$ in the MSSM or the type-II 2HDM.}
\end{figure}
%%%%%%%%%%%%%%%%%%%%%%%%%%%%%%%%%%%%%%%%%%%%%%%%%%%%%%%%%%%%%%

Note that $y^{2}_{e} \ll y^{2}_{\mu} \ll y^2_\tau \lesssim
0.25$ holds at the Fermi scale $\Lambda^{}_{\rm F}$ for $\tan\beta
\lesssim 50$, and their values decrease as the energy scale
grows up \cite{XZZ}. It is therefore an excellent approximation to take
$T^{}_l \simeq {\bf 1} - {\rm Diag}\{0, 0, \Delta^{}_\tau\}$ with
$\bf 1$ being the $3 \times 3$ unitary matrix and
\begin{eqnarray}
\Delta^{}_\tau = \frac{C^{}_l}{16\pi^{2}}
\int^{0}_{t^\prime} y^{2}_\tau \ {\rm d} t \; ,
%     (27)
\end{eqnarray}
which is a small quantity of ${\cal O}(0.1)$ or much smaller. To illustrate,
Figure 1 shows the numerical changes of $I^{}_G$ and $\Delta^{}_\tau$ with the energy
scale $\Lambda$ in the MSSM and the type-II 2HDM by fixing $\Lambda^{}_{\mu\tau} = 10^{14}$
GeV as the initial point and taking $\tan\beta = 10$ and
$30$ as two typical inputs. One can see that the signs of $\Delta^{}_\tau$
are opposite in these two scenarios, and thus they are distinguishable
at low energies. Now let us assume that the $\mu$-$\tau$ reflection symmetry of
$M^{}_\nu$ in Eq. (9) or $H^{}_\nu$ in Eq. (11) is realized
at $\Lambda^{}_{\mu\tau}$. Then at the electroweak scale $\Lambda^{}_{\rm F}$
we have
\begin{eqnarray}
H^{\prime}_\nu \simeq I^{2}_G \left[ H^{}_\nu - \Delta^{}_{\tau}
\left( \begin{matrix}0 & 0 & B^{\ast} \cr 0 & 0 & D \cr B & D^{\ast}
& 2C \end{matrix} \right) \right] \; ,
%  (28)
\end{eqnarray}
or equivalently,
\begin{eqnarray}
M^{\prime}_\nu \simeq I^{}_G \left[ M^{}_\nu -
\Delta^{}_{\tau} \left( \begin{matrix}0 & 0 & 0 \cr 0 & 0 & 0 \cr
e^{\ast} & \hspace{0.08cm} d^\ast \hspace{0.08cm}
& c^\ast \end{matrix} \right) \right] \; ,
%  (29)
\end{eqnarray}
in which the smallness of $\Delta^{}_\tau$ has been taken into account.
It is clear that the term proportional to $\Delta^{}_{\tau}$ measures
the strength of $\mu$-$\tau$ symmetry breaking. Even if
$M^{}_\nu$ is taken to be Hermitian, the
RGE-induced quantum correction will violate that Hermiticity at
$\Lambda < \Lambda^{}_{\mu\tau}$. In comparison, the Hermiticity of $H^{}_\nu$
is preserved in the whole RGE evolution from $\Lambda^{}_{\mu\tau}$ down
to $\Lambda^{}_{\rm F}$.

At this point it is worth comparing the generic expression of $H^\prime_\nu$
in Eq. (19) with the explicit one in Eq. (28). Of course, it is straightforward
to decompose the latter into a part respecting the $\mu$-$\tau$
reflection symmetry and a part violating this flavor symmetry, from which
one can easily obtain the dimensionless perturbation parameters
\begin{eqnarray}
\epsilon^{}_1 \simeq \frac{1}{2} \Delta^{}_\tau \; , \quad \quad
\epsilon^{}_2 \simeq \Delta^{}_\tau \; ,
%     (30)
\end{eqnarray}
implying that the only source of $\mu$-$\tau$ reflection symmetry breaking
in our example is the RGE-induced $\Delta^{}_\tau$ term. In practice,
it should be more convenient to directly use Eq. (28) to do a perturbation
calculation of the neutrino masses and flavor mixing parameters.

Before we start from Eq. (28) to derive the analytical expressions of
three neutrino masses and four flavor mixing parameters at
$\Lambda^{}_{\rm F}$ in the next section,
let us first derive two interesting relations with no need of doing
any perturbation calculation.
Eq. (11) tells us that $H^{}_\nu$ and
$H^{\prime}_\nu$ can be diagonalized by the unitary
matrices $U$ and $U^\prime$, respectively.
So the determinants of $H^{}_\nu$ and
$H^{\prime}_\nu$ are proportional to each other, giving rise to
\begin{eqnarray}
m^{\prime}_1 m^{\prime}_2 m^{\prime}_3 =
I^{3}_G I^{}_e I^{}_\mu I^{}_\tau \ m^{}_1 m^{}_2 m^{}_3 \; ,
%  (31)
\end{eqnarray}
with $m^{}_i$ and $m^{\prime}_i$ (for $i=1,2,3$) stand for the neutrino masses
at $\Lambda^{}_{\mu\tau}$ and $\Lambda^{}_{\rm F}$, respectively. Considering
the traces of $H^{}_\nu$ and $H^\prime_\nu$ in Eq. (24), we obtain
\begin{eqnarray}
\sum_{i} m^{\prime 2}_i = I^2_G \sum_\alpha I^2_\alpha \sum_i m^2_i
|V^{}_{\alpha i}|^2 \;
%     (32)
\end{eqnarray}
with $\alpha$ and $i$ running over $(e, \mu, \tau)$ and $(1, 2, 3)$, respectively.

But it is more interesting to establish an instructive relationship
between the Jarlskog invariant of CP violation ${\cal J}$
at $\Lambda^{}_{\mu\tau}$, defined through \cite{J}
\begin{eqnarray}
{\rm Im} \left(V^{}_{\alpha i} V^{}_{\beta j} V^*_{\alpha j}
V^*_{\beta i}\right) = {\cal J} \sum_\gamma \epsilon^{}_{\alpha\beta\gamma}
\sum_k \epsilon^{}_{ijk} \;
%     (33)
\end{eqnarray}
with the subscripts $(\alpha, \beta, \gamma)$ and $(i, j, k)$ running
respectively over $(e, \mu, \tau)$ and $(1, 2, 3)$,
and its counterpart ${\cal J}^\prime$ at
$\Lambda^{}_{\rm F}$. To do so, we first write out the elements of
$H^\prime_\nu$ in Eq. (24) in terms of the neutrino masses and the
PMNS matrix elements:
\begin{eqnarray}
\sum_i m^{\prime 2}_i U^{\prime}_{\alpha i} U^{\prime *}_{\beta i}
= I^2_G I^{}_\alpha I^{}_\beta \sum_i m^2_i U^{}_{\alpha i}
U^*_{\beta i} \; ,
%     (34)
\end{eqnarray}
in which both $\alpha$ and $\beta$ run over $e$, $\mu$ and $\tau$.
Note that $U = PV$ (or $U^\prime = P^\prime V^\prime$) contains three
unphysical phases. To eliminate them,
let us focus on the following rephasing invariant \cite{IJMPA}:
\begin{eqnarray}
\hspace{-0.2cm} & & \hspace{-0.2cm}
{\rm Im}\left[ \sum_i m^2_i U^{}_{e i} U^*_{\mu i} \cdot
\sum_j m^2_j U^{}_{\mu j} U^*_{\tau j} \cdot
\sum_k m^2_k U^{}_{\tau k} U^*_{e k} \right] \hspace{1cm}
\nonumber \\
\hspace{-0.2cm} & = & \hspace{-0.2cm}
\sum_i \sum_j \sum_k m^2_i m^2_j m^2_k \ {\rm Im} \left(
V^{}_{e i} V^{}_{\mu j} V^{}_{\tau k} V^*_{e k} V^*_{\mu i}
V^*_{\tau j}\right)
\nonumber \\
\hspace{-0.2cm} & = & \hspace{-0.2cm}
{\cal J} \sum_i \sum_j m^2_i m^4_j \sum_k \epsilon^{}_{ijk}
\nonumber \\
\hspace{-0.2cm} & = & \hspace{-0.2cm}
{\cal J} \Delta m^2_{21} \ \Delta m^2_{31} \ \Delta m^2_{32} \; ,
%     (35)
\end{eqnarray}
where the three neutrino mass-squared differences are defined
as $\Delta m^2_{ij} \equiv m^2_i - m^2_j$ (for $i,j = 1,2,3$).
Applying Eq. (34) to Eq. (35), we are then left with the elegant result
\begin{eqnarray}
{\cal J}^\prime \Delta m^{\prime 2}_{21} \Delta m^{\prime 2}_{31}
\Delta m^{\prime 2}_{32} = I^6_G I^2_e I^2_\mu I^2_\tau {\cal J}
\Delta m^2_{21} \Delta m^2_{31} \Delta m^2_{32} \; ,
%     (36)
\end{eqnarray}
which concisely connects the strength of leptonic CP violation at
$\Lambda^{}_{\mu\tau}$ to that at $\Lambda^{}_{\rm F}$.
Given the parametrization of $V$ in Eq. (13), the Jarlskog
invariant ${\cal J}$ reads as
\begin{eqnarray}
{\cal J} = \frac{1}{8} \sin 2\theta^{}_{12} \sin 2\theta^{}_{13}
\cos \theta^{}_{13} \sin 2\theta^{}_{23} \sin\delta \; .
%     (37)
\end{eqnarray}
If $\theta^{}_{23} = \pi/4$ and $\delta = \pi/2$ or $3\pi/2$
are taken into account in the $\mu$-$\tau$ reflection symmetry
limit, then we arrive at $|{\cal J}| = \sin 2\theta^{}_{12}
\sin 2\theta^{}_{13} \cos \theta^{}_{13}/8$. Taking a similar
parametrization for $V^\prime$, one may express ${\cal J}^\prime$
in terms of the corresponding flavor mixing parameters as
\begin{eqnarray}
{\cal J}^\prime = \frac{1}{8} \sin 2\theta^{\prime}_{12}
\sin 2\theta^{\prime}_{13} \cos \theta^{\prime}_{13}
\sin 2\theta^{\prime}_{23} \sin\delta^\prime \; .
%     (38)
\end{eqnarray}
In the next section we shall establish
the analytical relations between $(\theta^{}_{12}, \theta^{}_{13},
\theta^{}_{23}, \delta)$ at $\Lambda^{}_{\mu\tau}$ and
$(\theta^{\prime}_{12}, \theta^{\prime}_{13}, \theta^{\prime}_{23},
\delta^\prime)$ at $\Lambda^{}_{\rm F}$ in a perturbation approach.

\section{RGE corrections to $U$}

Let us start from Eq. (28) to do a perturbation calculation in
order to derive the analytical expressions of three neutrino
masses and four flavor mixing parameters at
$\Lambda^{}_{\rm F}$. Similar to $H^{}_\nu$ in Eq. (11), $H^\prime_\nu$
can also be reconstructed in the same way:
\begin{eqnarray}
H_{\nu}^{\prime} \equiv M^\prime_\nu M^{\prime \dagger}_\nu
= U^{\prime} \hat M_{\nu}^{\prime 2} U^{\prime\dagger} \; ,
%     (39)
\end{eqnarray}
in which $U^\prime = P^\prime V^\prime$ with $P^\prime$ being a
diagonal phase matrix, and $\hat M_{\nu}^{\prime} \equiv {\rm Diag}
\{m_1^{\prime}, m_2^{\prime}, m_3^{\prime}\}$ with $m^\prime_i$
being the neutrino masses at
$\Lambda^{}_{\rm F}$. Then the approximate relationship between
$H^\prime_\nu$ and $H^{}_\nu$ in Eq. (28) can be rewritten as
\begin{eqnarray}
\hat M^{\prime 2 }_\nu \simeq I^{2}_G U^{\prime\dagger}
\left[ U \hat M^{2}_\nu U^{\dagger} -
\Delta^{}_{\tau}\left( \begin{matrix}0 & 0 & \sum\limits_{i}^{}
m_i^{2} U_{e i}^{} U_{\tau i}^{\ast} \cr 0 & 0 &
\sum\limits_{i}^{} m_i^{2} U_{\mu i}^{} U_{\tau i}^{\ast} \cr
\sum\limits_{i}^{} m_i^{2} U_{e i}^{\ast} U_{\tau i}^{} &
\sum\limits_{i}^{} m_i^{2} U_{\mu i}^{\ast} U_{\tau i}^{}
& 2\sum\limits_{i}^{} m_i^{2} |U_{\tau i}^{} |^2 \end{matrix}
\right) \right]U^{\prime}_{} \; .
%  (40)
\end{eqnarray}
Treating $\Delta^{}_{\tau}$ as a small perturbation parameter, let us
define the RGE-induced deviations of the relevant flavor mixing angles
and phase parameters at $\Lambda^{}_{\rm F}$ from their original
counterparts at $\Lambda^{}_{\mu\tau}$ as follows:
\begin{eqnarray}
\Delta \theta^{}_{12} \hspace{-0.2cm} & = & \hspace{-0.2cm}
\theta^{\prime}_{12} - \theta^{}_{12} \; , \quad\quad
\Delta \delta = \delta^{\prime} - \delta \; ,
\nonumber \\
\Delta \theta^{}_{13} \hspace{-0.2cm} & = & \hspace{-0.2cm}
\theta^{\prime}_{13} - \theta^{}_{13} \; , \quad\quad
\Delta\phi^{}_{e\mu} = (\phi^{\prime}_{e} - \phi^{\prime}_{\mu})
- (\phi^{}_{e} - \phi^{}_{\mu}) \; ,
\nonumber \\
\Delta \theta^{}_{23} \hspace{-0.2cm} & = & \hspace{-0.2cm}
\theta^{\prime}_{23} - \theta^{}_{23} \; , \quad\quad
\Delta \phi^{}_{e\tau} = (\phi^{\prime}_{e} - \phi^{\prime}_{\tau})
- (\phi^{}_{e} - \phi^{}_{\tau}) \; ,
%   (41)
\end{eqnarray}
which are expected to be small enough in magnitude as
compared with their respective starting values at $\Lambda^{}_{\mu\tau}$.
Note that $\theta^{}_{23} = \pi/4$ and $\delta = \pi/2$ or
$3\pi/2$ at the $\mu$-$\tau$ reflection symmetry scale
$\Lambda^{}_{\mu\tau}$ will be implied in the
subsequent perturbation calculations.
Note also that only two combinations of the three unphysical phases
in $P$ or $P^\prime$, as indicated in Eq. (41), are associated with our
derivation of the RGEs for the physical parameters. They ought
not to be ignored in the course of the calculations, but of
course they do not show up in the final results of
$\Delta\theta^{}_{12}$, $\Delta\theta^{}_{13}$, $\Delta\theta^{}_{23}$ and
$\Delta\delta$. Next we expand the elements of $\hat M_{\nu}^{\prime 2}$
in terms of the above perturbation parameters and only keep their
first-order contributions.

First of all, it is straightforward to
obtain the analytical results of three neutrino masses from the
diagonal elements of $\hat M_{\nu}^{\prime 2}$. Namely,
\begin{eqnarray}
m^{\prime}_{1} \hspace{-0.2cm} & \simeq & \hspace{-0.2cm}
I^{}_{G} m^{}_{1} \left[1 - \frac{1}{2} \Delta^{}_{\tau}
\left(s^{2}_{12} c^{2}_{13} + s^{2}_{13}\right)\right] \; ,
\nonumber \\
m^{\prime}_{2} \hspace{-0.2cm} & \simeq & \hspace{-0.2cm}
I^{}_{G} m^{}_{2} \left[1 - \frac{1}{2}\Delta^{}_{\tau}
\left(c^{2}_{12} c^{2}_{13} + s^{2}_{13}\right)\right] \; ,
\nonumber \\
m^{\prime}_{3} \hspace{-0.2cm} & \simeq & \hspace{-0.2cm}
I^{}_{G} m^{}_{3} \left[1 - \frac{1}{2}\Delta^{}_{\tau}
c^{2}_{13}\right] \; .
%   (42)
\end{eqnarray}
Obviously but interestingly, $m^\prime_i / m^{}_i \simeq I^{}_G$
holds in the leading-order approximation, implying that the
three neutrino masses almost run in step.
Given $I^{}_e \simeq I^{}_\mu \simeq 1$ and $I^{}_\tau \simeq
1 - \Delta^{}_\tau$ and the $\mu$-$\tau$ reflection symmetry
at $\Lambda^{}_{\mu\tau}$,
it is easy to check that the product of
$m^\prime_1$, $m^\prime_2$ and $m^\prime_3$ in Eq. (42) can
successfully reproduce the elegant relationship achieved in Eq. (31).
Moreover, Eq. (42) leads us to the sum rule
\begin{eqnarray}
\sum\limits_i {m^{\prime 2}_{i}} \simeq I^{2}_{G} \sum\limits_i{m^{2}_{i}
\left(1 - 2\Delta^{}_{\tau} |V^{}_{\tau i}|^2 \right)} \; ,
%     (43)
\end{eqnarray}
which is consistent with the more generic one derived in
Eq. (32) if the same approximations are made and the $\mu$-$\tau$
reflection symmetry is taken into account.

Second, the off-diagonal elements of $\hat M_{\nu}^{\prime 2}$ in
Eq. (40) must vanish, yielding the following six constraint equations
in our analytical approximations:
\begin{eqnarray}
&& 2 \Delta m_{21}^{2} \Delta \theta_{12}^{} +
\eta  s_{13}^{}  \Delta m_{21}^{2} \left(\Delta \phi_{e\mu}^{} -
\Delta \phi_{e\tau}^{}\right) - c_{12}^{} s_{12}^{} c_{13}^2
\overline m_{12}^{} \Delta_{\tau}^{} \simeq 0 \; ,
\nonumber \\
&& 2 \left(c_{12}^{2} - s_{12}^{2}\right)
s_{13}^{} \Delta m_{21}^2 \Delta \theta_{23}^{}
- \eta c_{12}^{} s_{12}^{} \Delta m_{21}^2 \left[ 2 s_{13}^2
\Delta \delta + c_{13}^2 \left(\Delta \phi_{e\mu}^{} + \Delta
\phi_{e\tau}^{}\right) \right] + s_{13}^{} \overline m_{12}^{}
\Delta^{}_{\tau} \simeq 0 \; ,
\nonumber \\
&& 2 s_{12}^{} c_{13}^{} \Delta m_{31}^{2}
\Delta \theta_{23}^{} + \eta c_{12}^{} c_{13}^{} s_{13}^{}
\Delta m_{31}^{2} \left(2 \Delta \delta -\Delta \phi_{e\mu}^{} -
\Delta \phi_{e\tau}^{}\right) - s_{12}^{} c_{13}^{} \overline m_{13}^{}
\Delta_{\tau}^{} \simeq 0 \; ,
\nonumber \\
&& 2 c_{12}^{} \Delta m_{31}^2 \Delta
\theta_{13}^{} - \eta s_{12}^{} c_{13}^{} \Delta m^2_{31} \left(\Delta
\phi_{e\mu}^{} -\Delta \phi_{e\tau}^{}\right) - c_{12}^{} c_{13}^{}
s_{13}^{} \overline m_{13}^{} \Delta_{\tau}^{} \simeq 0 \; ,
\nonumber \\
&& 2 c_{12}^{} c_{13}^{} \Delta m_{32}^2 \Delta
\theta_{23}^{} -\eta s_{12}^{} c_{13}^{} s_{13}^{} \Delta
m_{32}^2 \left(2 \Delta \delta - \Delta \phi_{e\mu}^{} -
\Delta \phi_{e\tau}^{}\right) - c_{12}^{} c_{13}^{}
\overline m_{23}^{} \Delta_{\tau}^{} \simeq 0 \; ,
\nonumber \\
&& 2 s_{12}^{} \Delta m_{32}^{2}
\Delta \theta_{13}^{} + \eta c_{12}^{} c_{13}^{} \Delta m_{32}^{2}
\left(\Delta \phi_{e\mu}^{} - \Delta \phi_{e\tau}^{}\right) -
s_{12}^{} c_{13}^{} s_{13}^{} \overline m_{23}^{} \Delta_{\tau}^{}
\simeq 0 \; ,
%     (44)
\end{eqnarray}
where $\eta \equiv \sin\delta = \pm 1$ in the $\mu$-$\tau$
reflection symmetry limit, and
$\overline m_{ij}^{} \equiv m_i^{2} + m_j^2$ (for $i, j=1,2,3$).
Solving the above equations, we obtain
\begin{eqnarray}
\Delta\theta^{}_{12} \hspace{-0.2cm} & \simeq & \hspace{-0.2cm}
\Delta^{}_{\tau} s^{}_{12}c^{}_{12}
\left[ \frac{m^{2}_{1} + m^{2}_{2}}{2 \Delta m^{2}_{21}} c^{2}_{13} -
\frac{m^{2}_{3}\Delta m^{2}_{21}}{\Delta m^{2}_{31}
\Delta m^{2}_{32}} s^{2}_{13} \right] \; ,
\nonumber \\
\Delta\theta^{}_{13} \hspace{-0.2cm} & \simeq & \hspace{-0.2cm}
\Delta^{}_{\tau} s^{}_{13}c^{}_{13}
\left[ \frac{m^{2}_{2} + m^{2}_{3}}{2 \Delta m^{2}_{32}} s^{2}_{12} +
\frac{m^{2}_{1} + m^{2}_{3}}{2\Delta m^{2}_{31}} c^{2}_{12} \right] \; ,
\nonumber \\
\Delta\theta^{}_{23} \hspace{-0.2cm} & \simeq & \hspace{-0.2cm}
\Delta^{}_{\tau} \left[ \frac{m^{2}_{2} + m^{2}_{3}}
{2 \Delta m^{2}_{32}} c^{2}_{12} +\frac{m^{2}_{1} + m^{2}_{3}}{2\Delta
m^{2}_{31}} s^{2}_{12} \right] \; ;
%   (45)
\end{eqnarray}
and
\begin{eqnarray}
&& \Delta\delta^{}  \simeq
\eta \Delta^{}_{\tau} \left[ \frac{m^{2}_{1} \Delta m^{2}_{32}}
{\Delta m^{2}_{21} \Delta m^{2}_{31}} t^{}_{12}s^{}_{13} +
\frac{m^{2}_{2} \Delta m^{2}_{31}}{\Delta m^{2}_{21}
	\Delta m^{2}_{32}} \cdot \frac{s^{}_{13}}{t^{}_{12}} - \frac{m^{2}_{3}
	\Delta m^{2}_{21}}{\Delta m^{2}_{31} \Delta m^{2}_{32}} s^{}_{12}c^{}_{12}
\left( \frac{1}{s^{}_{13}} + s^{}_{13} \right) \right] \; ,
\nonumber \\
&& \Delta\phi^{}_{e\mu}  \simeq
\eta \Delta_\tau s^{}_{13} \left[\frac{m^2_1 \Delta m^2_{32}} {\Delta m^2_{21}
	\Delta m^2_{31}} t^{}_{12} + \frac{m^2_2 \Delta m^2_{31}}{\Delta m^2_{21}
	\Delta m^2_{32} } \cdot \frac{1}{t^{}_{12}}  \right] \;,
\nonumber \\
&& \Delta\phi^{}_{e\tau}  \simeq
\eta \Delta_\tau s^{}_{13} \left[ \frac{m^2_1 \Delta m^2_{32}} {\Delta m^2_{21}
	\Delta m^2_{31}} t^{}_{12} +\frac{m^2_2 \Delta m^2_{31}}{\Delta m^2_{21}
	\Delta m^2_{32} } \cdot \frac{1}{t^{}_{12}} - \frac{2 m^2_3 \Delta m^2_{21}}
{\Delta m^2_{31} \Delta m^2_{32}} s^{}_{12} c^{}_{12} \right] \; ,
%		(46)
\end{eqnarray}
where $t^{}_{12} \equiv \tan \theta_{12}^{}$. One can see that the RGE-induced
corrections to all the four flavor mixing parameters are proportional to
$\Delta^{}_\tau$, a fact which is under rational expectation. Among the
three angles, $\theta^{}_{12}$ is more sensitive to the quantum
corrections than $\theta^{}_{13}$ and $\theta^{}_{23}$ in most cases,
mainly because of
$|\Delta m^2_{31}| \simeq |\Delta m^2_{32}| \sim 30 \Delta m^2_{21}$ \cite{FIT}.
On the other hand, the smallness of $s^{}_{13}$ \cite{XZ2012} implies that the magnitude of
$\Delta\theta^{}_{13}$ must be smaller than that of $\Delta\theta^{}_{23}$.
But the expression of $\Delta\delta$ contains three terms proportional
to $s^{}_{13}$ and one term proportional to $1/s^{}_{13}$, and hence
the overall running effect of $\delta$ is generally expected to be more
significant than those of three flavor mixing angles, or at least than
those of $\theta^{}_{13}$ and $\theta^{}_{23}$. Note that
$\phi^{}_\mu + \phi^{}_\tau = 2 \phi^{}_e$ holds at $\Lambda^{}_{\mu\tau}$
due to the $\mu$-$\tau$ reflection symmetry of $H^{}_\nu$, and hence
$2\phi^{\prime}_e - \phi^\prime_\mu - \phi^\prime_\tau =
\Delta \phi^{}_{e\mu} + \Delta \phi^{}_{e\tau} \propto \Delta^{}_{\tau}$
is not vanishing at $\Lambda^{}_{\rm F}$, providing us with another
(unphysical) measure of the RGE-induced $\mu$-$\tau$ reflection symmetry
breaking of $H^\prime_\nu$.

There are two ways to calculate the Jarlskog invariant ${\cal J}^\prime$
at $\Lambda^{}_{\rm F}$: one is to apply Eq. (42) to the elegant
relationship between ${\cal J}$ and ${\cal J}^\prime$ in Eq. (36) with
$I^{}_e \simeq I^{}_\mu \simeq 1$ and $I^{}_\tau \simeq
1 - \Delta^{}_\tau$, and the other is to do a direct perturbation
calculation of ${\cal J}^\prime$ by using Eqs. (38), (45) and (46).
After doing such a calculation, we obtain the ratio of ${\cal J}^\prime$
at $\Lambda^{}_{\rm F}$ to ${\cal J}$ at $\Lambda^{}_{\mu\tau}$ as
follows:
\begin{eqnarray}
\frac{{\cal J}^{\prime}}{\cal J} \simeq 1 + \Delta^{}_{\tau} \left[
\left( s^{2}_{12}c^{2}_{13} - s^{2}_{13}\right) \left( \frac{m^{2}_{2}}
{\Delta m^{2}_{32}} - \frac{m^{2}_{1}}{\Delta m^{2}_{21}} \right) +
\left( c^{2}_{12}c^{2}_{13} - s^{2}_{13}\right) \left( \frac{m^{2}_{1}}
{\Delta m^{2}_{31}} + \frac{m^{2}_{2}}{\Delta m^{2}_{21}} \right) \right]
\; .
%    (47)
\end{eqnarray}
Different from $\delta$, ${\cal J}$ evolves in a way insensitive to
the smallness of $\theta^{}_{13}$.

We proceed to numerically illustrate the RGE-induced corrections
to the neutrino masses and flavor mixing parameters in the MSSM 
and the type-II 2HDM by using the program advocated in 
Ref. \cite{Program} and taking
$\Lambda^{}_{\mu\tau} = 10^{14}$ GeV as a typical choice, where
$\theta^{}_{23} = \pi/4$ and $\delta = 3\pi/2$
are input. For the sake of simplicity, we adjust the initial
values of $m^{}_1$ (or $m^{}_3$), $\Delta m^2_{21}$,
$\Delta m^2_{31}$, $\theta^{}_{12}$ and $\theta^{}_{13}$ to make sure
that all the neutrino oscillation parameters can be compatible
with current experimental data at $\Lambda^{}_{\rm F}$ \cite{FIT}.
The main numerical results are summarized in Tables 1 and 2 as well as
Figures 2, 3 and 4, in which two possibilities of the neutrino mass
spectrum have been taken into account --- the normal hierarchy (NH) with
$m^{}_1 < m^{}_2 < m^{}_3$ or $\Delta m^2_{31} >0$ and the
inverted hierarchy (IH) with $m^{}_3 < m^{}_1 < m^{}_2$ or
$\Delta m^2_{31} <0$. Some comments and discussions are in order.
%%%%%%%%%%%%%%%%%%%%%%%%%%%%%%%%%%%%%%%%%%%%%%%%%%%%%%%%%%%%%%
%%%%%%%%%%%%%% Table 1 %%%%%%%%%%%%%%%%%%%%%%%%%%%%%%%%%%%%%%%
\begin{table}[h!]\centering
\caption{An illustration of the neutrino oscillation parameters
at $\Lambda^{}_{\mu\tau}$ and $\Lambda^{}_{\rm F}$ in the MSSM
with $\tan{\beta} =10$ or $30$, where both NH and IH cases are
considered.}
\vspace{0.25cm}
	\renewcommand\arraystretch{1.3}
	\begin{tabular}{lllllllll}
		\hline\hline
		MSSM & \multicolumn{2}{l}{NH, $\tan{\beta}=10$} &
		\multicolumn{2}{l}{NH, $\tan{\beta}=30$} & \multicolumn{2}{l}{IH, $\tan{\beta}=10$} &
		\multicolumn{2}{l}{IH, $\tan{\beta}=30$}
		\\
		Parameter & $\Lambda^{}_{\mu\tau}$ & $\Lambda^{}_{\rm F}$ & $\Lambda^{}_{\mu\tau}$ &
		$\Lambda^{}_{\rm F}$ & $\Lambda^{}_{\mu\tau}$ & $\Lambda^{}_{\rm F}$ &
		$\Lambda^{}_{\mu\tau}$ & $\Lambda^{}_{\rm F}$
		\\
		\hline
		$m^{}_{\rm lightest} \;[10^{-2} ~\rm eV]$ & 5.4 & 5.06 & 5.4 & 5.03 & 5.36 & 5.02 & 5.4 & 5.00
		\\
		$ \Delta m^{2}_{21} \;[10^{-5} ~{\rm eV^{2}}]$ & 8.77 & 7.56 & 10.77 & 7.56 & 8.92 & 7.56 & 13.19 & 7.56
		\\
		$ |\Delta m^{2}_{31}| \;[10^{-3} ~{\rm eV^{2}}]$ & 2.91 & 2.55 & 3.00 & 2.55 & 2.84 & 2.49 & 2.84 & 2.49
		\\
		$ \theta^{}_{12} \;[ ^{\circ}]$ & 33.31 & 34.50 & 24.36 & 34.51 & 32.23 & 34.50 & 18.30 & 34.49
		\\
		$ \theta^{}_{13} \;[ ^{\circ}]$ & 8.42 & 8.44 & 8.28 & 8.44 & 8.43 & 8.41 & 8.58 & 8.41
		\\
		$ \theta^{}_{23} \;[ ^{\circ}]$ & 45 & 45.12 & 45 & 46.17 & 45 & 44.89 & 45 & 43.87
		\\
		$ \delta \;[ ^{\circ}]$ & 270 & 269.18 & 270 & 261.87 & 270 & 268.42 & 270 & 254.15
		\\
		\hline
		${\cal J} \;[10^{-2}] $ & $-3.29$ & $-3.35$ & $-2.65$ & $-3.32$ & $-3.24$ & $-3.34$ & $-2.17$ & $-3.21$
		\\
		\hline
		\hline
	\end{tabular}
\end{table}
%%%%%%%%%%%%%%%%%%%%%%%%%%%%%%%%%%%%%%%%%%%%%%%%%%%%%%%%%%%%%%
\vspace{0.2cm}
%%%%%%%%%%%%%% Table 2 %%%%%%%%%%%%%%%%%%%%%%%%%%%%%%%%%%%%%%%
\begin{table}[h!]\centering
	\caption{An illustration of the neutrino oscillation parameters
		at $\Lambda^{}_{\mu\tau}$ and $\Lambda^{}_{\rm F}$ in the type-II 2HDM
		with $\tan{\beta} =10$ or $30$, where both NH and IH cases are
		considered.}
	\vspace{0.25cm}
	\renewcommand\arraystretch{1.3}
	\begin{tabular}{lllllllll}
		\hline\hline
		2HDM & \multicolumn{2}{l}{NH, $\tan{\beta}=10$} &
		\multicolumn{2}{l}{NH, $\tan{\beta}=30$} & \multicolumn{2}{l}{IH, $\tan{\beta}=10$} &
		\multicolumn{2}{l}{IH, $\tan{\beta}=30$}
		\\
		Parameter & $\Lambda^{}_{\mu\tau}$ & $\Lambda^{}_{\rm F}$ & $\Lambda^{}_{\mu\tau}$ &
		$\Lambda^{}_{\rm F}$ & $\Lambda^{}_{\mu\tau}$ & $\Lambda^{}_{\rm F}$ &
		$\Lambda^{}_{\mu\tau}$ & $\Lambda^{}_{\rm F}$
		\\
		\hline
		$m^{}_{\rm lightest} \;[10^{-2} ~\rm eV]$ & 4.29 & 5.0 & 4.53 & 5.01 & 4.29 & 5.0 & 4.5 & 5.0
		\\
		$ \Delta m^{2}_{21} \;[10^{-5} ~{\rm eV^{2}}]$ & 5.44 & 7.56 & 5.94 & 7.56 & 5.34 & 7.56 & 8.31 & 7.56
		\\
		$ |\Delta m^{2}_{31}| \;[10^{-3} ~{\rm eV^{2}}]$ & 1.88 & 2.55 & 2.04 & 2.55 & 1.84 & 2.49 & 2.04 & 2.49
		\\
		$ \theta^{}_{12} \;[ ^{\circ}]$ & 36.43 & 34.50 & 54.36 & 34.55 & 38.45 & 34.52 & 70.6 & 34.52
		\\
		$ \theta^{}_{13} \;[ ^{\circ}]$ & 8.47 & 8.44 & 8.72 & 8.44 & 8.38 & 8.41 & 8.17 & 8.41
		\\
		$ \theta^{}_{23} \;[ ^{\circ}]$ & 45 & 44.81 & 45 & 43.18 & 45 & 45.18 & 45 & 46.77
		\\
		$ \delta \;[ ^{\circ}]$ & 270 & 271.29 & 270 & 282.71 & 270 & 272.52 & 270 & 295.34
		\\
		\hline
		${\cal J} \;[10^{-2}] $ & $-3.44$ & $-3.35$ & $-3.51$ & $-3.27$ & $-3.47$ & $-3.34$ & $-2.18$ & $-3.01$
		\\
		\hline
		\hline
	\end{tabular}
\end{table}
%%%%%%%%%%%%%%%%%%%%%%%%%%%%%%%%%%%%%%%%%%%%%%%%%%%%%%%%%%%%%%%%%%%
%%%%%%%%%%%%%%%%%%%%%%%%%%% Figure 2 %%%%%%%%%%%%%%%%%%%%%%%%%%%%%%
\begin{figure}[h!]
	\centering
\includegraphics[]{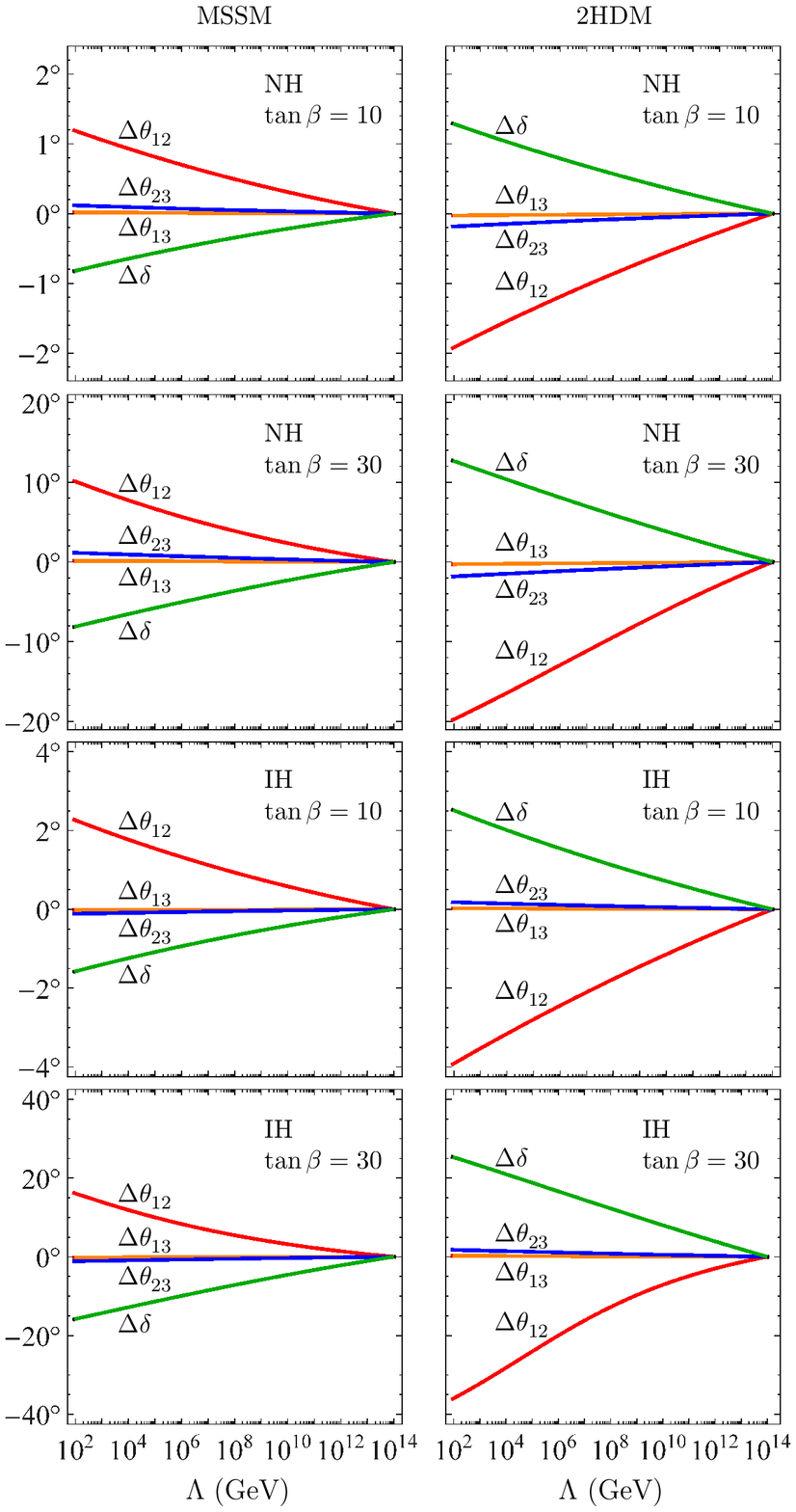}
\caption{The changes of $\Delta\theta^{}_{12}$, $\Delta\theta^{}_{13}$,
$\Delta\theta^{}_{23}$ and $\Delta\delta$ with the energy scale
$\Lambda$ in the MSSM and the type-II 2HDM, where $m^{}_{\rm lightest} =0.05 ~\rm eV$ 
at $\Lambda^{}_{\rm F} = 10^2 ~\rm GeV$ is typically input
and $\theta^{}_{23}=\pi/4$ and $\delta=3\pi/2$ at $\Lambda^{}_{\mu\tau}
= 10^{14} ~{\rm GeV}$ are fixed by the $\mu$-$\tau$ reflection symmetry.}
\end{figure}
%%%%%%%%%%%%%%%%%%%%%%%%%%%%%%%%%%%%%%%%%%%%%%%%%%%%%%%%%%%%%%%%%%%

(1) In the MSSM, Table 1 and Figure 2 
show that the values of three flavor mixing angles
increase in the NH case as the energy scale $\Lambda$ decreases, but
$\theta^{}_{13}$ and $\theta^{}_{23}$
decrease in the IH case as $\Lambda$ decreases. In either case a larger
value of $\tan\beta$ will enhance the running effects. Such a direction
of evolution of $\Delta\theta^{}_{ij}$ (for $ij =12, 13$ or $23$) can
easily be understood from our analytical approximations made in Eq. (45). 
In comparison,
the CP-violating phase $\delta$ decreases in both NH and IH cases
when $\Lambda$ becomes lower. The reason for this behavior
can be seen in Eq. (46) --- namely,
$\delta = 3\pi/2$ (or $\eta =-1$) has been input at $\Lambda^{}_{\mu\tau}$,
and $\Delta\delta$ is essentially insensitive to the sign of
$\Delta m^2_{31}$ which is always the same as the sign of $\Delta m^2_{32}$.
Moreover, both Table 1 and Figure 3 tell us that the magnitude of the
Jarlskog invariant (i.e., $|{\cal J}|$)
increases as $\Lambda$ decreases, no matter whether the neutrino mass
hierarchy is normal or inverted. Eq. (47) shows that the ratio
${\cal J}^\prime/{\cal J}$ must be slightly larger than one if the
term proportional to $m^2_2/\Delta m^2_{21}$ is dominant. Although
the above observations are more or less subject to the limited parameter
space that we have taken into account, our analytical results in
Eqs. (45), (46) and (47) are certainly more general and more useful.
%%%%%%%%%%%%%%%%%%%%%%%%%% Figure 3 %%%%%%%%%%%%%%%%%%%%%%%%%%%%%%%
\begin{figure}[h!]
\includegraphics[width=16.4cm]{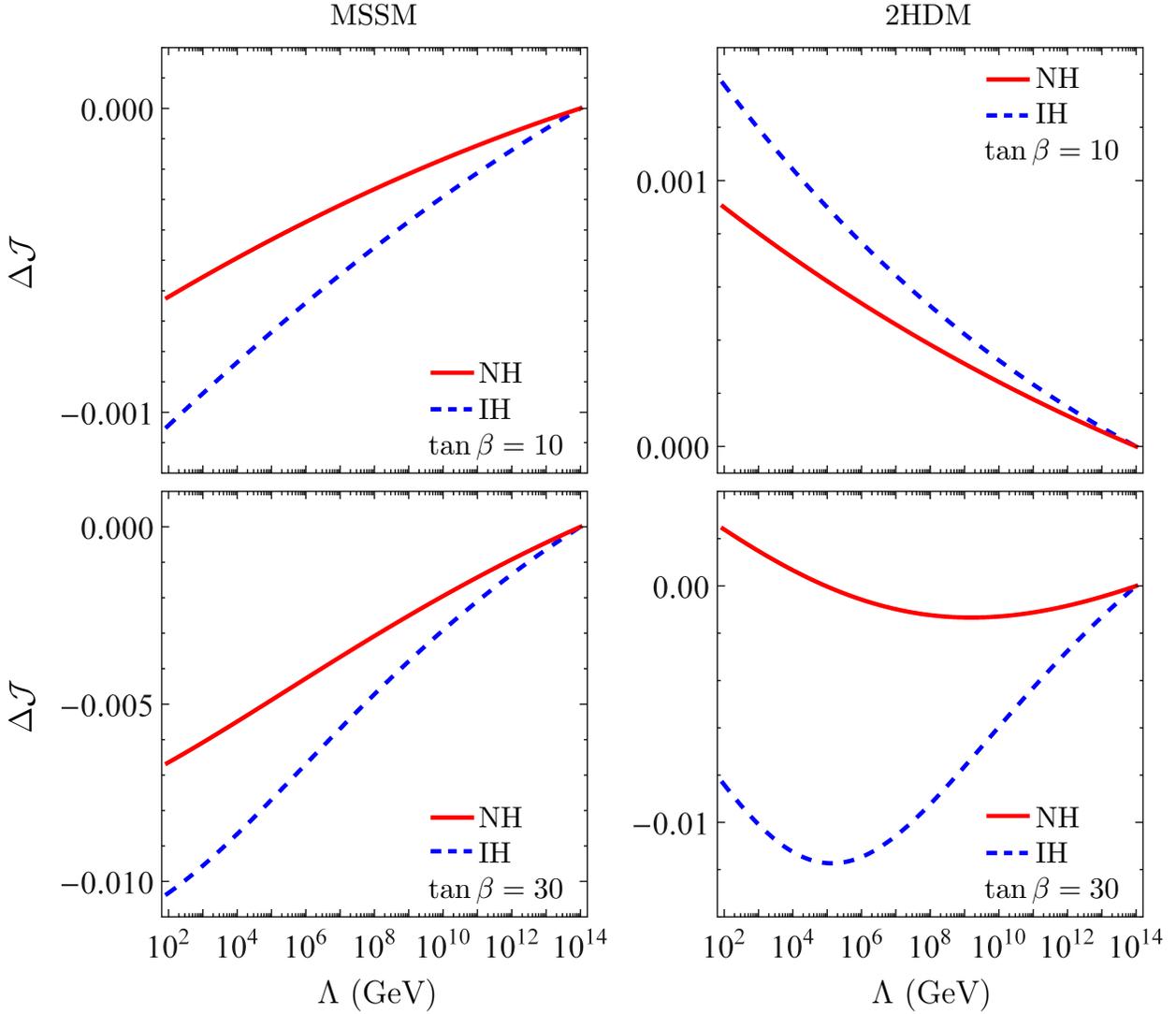}
\caption{An illustration of the change of $\Delta {\cal J} \equiv
{\cal J}^\prime - {\cal J}$ with the energy scale
$\Lambda$ in the MSSM and the type-II 2HDM, where $m^{}_{\rm lightest} =0.05 ~\rm eV$ 
at $\Lambda^{}_{\rm F} = 10^2 ~\rm GeV$ is typically input
and $\theta^{}_{23}=\pi/4$ and $\delta=3\pi/2$ at $\Lambda^{}_{\mu\tau}
= 10^{14} ~{\rm GeV}$ are fixed by the $\mu$-$\tau$ reflection symmetry.}
\end{figure}
%%%%%%%%%%%%%%%%%%%%%%%%%%%%%%%%%%%%%%%%%%%%%%%%%%%%%%%%%%%%%%%%%%%

(2) In the type-II 2HDM, the running behaviors of $\theta^{}_{12}$,
$\theta^{}_{13}$, $\theta^{}_{23}$ and $\delta$
take the opposite directions as compared
with those in the MSSM. The reason is simply that the signs of $\Delta^{}_\tau$
are opposite in these two scenarios. Because of $C^{}_l = 1$ in the
MSSM and $C^{}_l = -3/2$ in the type-II 2HDM, the magnitude $\Delta^{}_\tau$ in
the latter case is about 1.5 times larger than that in the former case.  
That is why we have taken the type-II 2HDM scenario for our numerical 
illustration, in contrast with the MSSM scenario. Note, however, that the
evolution of $\Delta {\cal J}$ with $\Lambda$ is a bit subtle in the 
type-II 2HDM case when $\tan\beta$ is sufficiently large. For example, 
the minimum of $\Delta {\cal J}$ shown in the right-bottom panel of Figure 3 
is expected to arise from a significant cancellation among the terms on
the right-hand side of Eq. (47). 

(3) It is worth highlighting that the RGE-induced effect of $\mu$-$\tau$
reflection symmetry breaking provides a model-independent way to connect
three burning issues in today's neutrino physics: the neutrino mass ordering,
the octant of $\theta^{}_{23}$ and leptonic CP violation. Some interesting
works have been done in this regard in the case that the massive neutrinos are
the Majorana particles \cite{XZ2016,Luo,Zhao,Xu}. Here we have discussed how the
$\mu$-$\tau$ reflection symmetry of Dirac neutrinos can be spontaneously
broken by the RGE evolution from $\Lambda^{}_{\mu\tau}$ down to
$\Lambda^{}_{\rm F}$ in the MSSM and the type-II 2HDM, 
and how this symmetry breaking affects
the octant of $\theta^{}_{23}$ and the quadrant of $\delta$ in both
NH and IH cases. As shown in Figure 2, the type-II 2HDM scenario 
seems to be somewhat favored 
if we stick to the best-fit value of $\theta^{}_{23}$ at low energy scales
\cite{FIT}, which lies in the first octant in the NH case but in the second
octant in the IH case
%%%%%%%%%%%%%%%%%%%%%%%%%%%%%%%%%%%
\footnote{If one works on the RGEs in the SM framework, then
$\Delta\theta^{}_{ij}$ and $\Delta\delta$ will evolve with the 
energy scales in a similar way as in the type-II 2HDM scenario. 
In this case, however, the running effects of relevant
parameters are expected to be much milder because of the lack of
the $\tan\beta$ enhancement. More seriously, the SM-like RGEs may
suffer from the vacuum-stability problem as the energy scale is above
$10^{10}$ GeV \cite{XZZ}.}.
%%%%%%%%%%%%%%%%%%%%%%%%%%%%%%%%%%%
For the time being, however, the ``best-fit" values of $\theta^{}_{23}$ 
from a global analysis of current 
neutrino oscillation data should not be taken too seriously, because their
statistical significance remains rather poor \cite{FIT}. It is more
appropriate to consider the $2\sigma$ or $3\sigma$ intervals of those
neutrino oscillation parameters, in which case the octant of
$\theta^{}_{23}$ is not yet fixed
%%%%%%%%%%%%%%%%%%%%%%%%%%%%%%%%%%%%%%%%%%%%%%%%%%%%%%%%%%%%%%%%%%%%%%%%%%%%%%
\footnote{At this point it is worth mentioning that 
the latest T2K neutrino oscillation result provides a very preliminary hint 
that $\theta^{}_{23}$ might lie in the second octant in the NH case \cite{T2K},
a possibility compatible with our results 
in the MSSM scenario shown in Table 1 and Figure 2.}.
%%%%%%%%%%%%%%%%%%%%%%%%%%%%%%%%%%%%%%%%%%%%%%%%%%%%%%%%%%%%%%%%%%%%%%%%%%%%%%

(4) As a by-product, Figure 4 illustrates the evolution behaviors
of three neutrino masses in both NH and IH cases. Since we have
intended to take $m^{}_{\rm lightest} =0.05 ~{\rm eV}$ at
$\Lambda^{}_{\rm F}$ in our numerical calculations so as to
reasonably magnify the RGE running effects, the neutrino mass
spectrum is not far away from the nearly degenerate case with
a fine split between $m^{}_1$ and $m^{}_2$ even if it is
normal. Our numerical results are consistent with the analytical
ones obtained in Eq. (42) --- namely, the evolution of $m^{}_i$
is mainly governed by that of $I^{}_G$ and thus insensitive to
the value of $\tan\beta$. For the same reason, the results of
$m^{}_i$ in the MSSM are not very different from those in the
type-II 2HDM.  
%%%%%%%%%%%%%%%%%%%%%%%%% Figure 4 %%%%%%%%%%%%%%%%%%%%%%%%%%%%%%%%
\begin{figure}[h!]
	\centering
\includegraphics[]{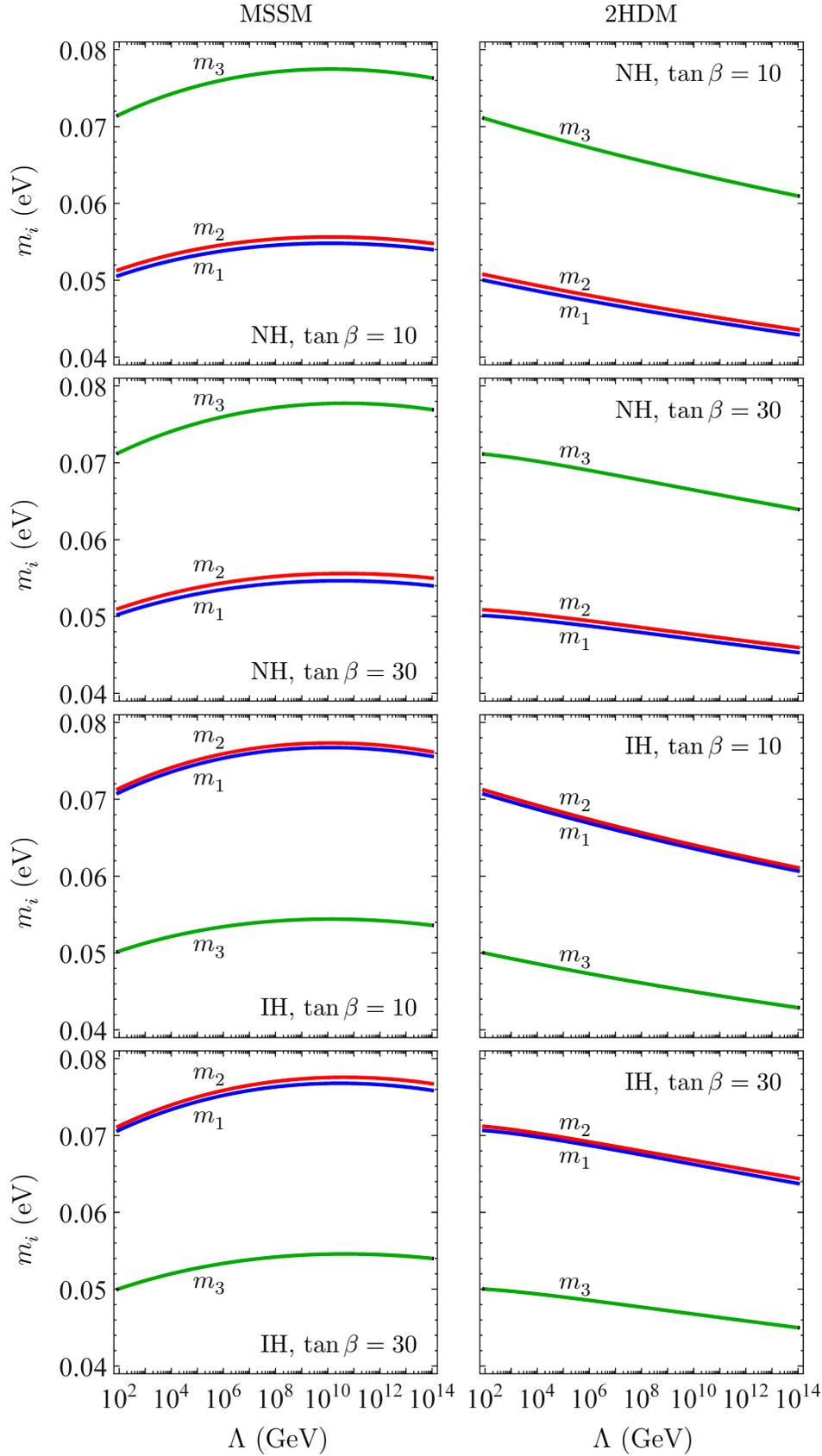}
\caption{The three neutrino masses evolving with the energy scale
$\Lambda$ in the MSSM and the type-II 2HDM, where $m^{}_{\rm lightest} =0.05 ~\rm eV$
at $\Lambda^{}_{\rm F} = 10^2 ~\rm GeV$ is typically input
and $\theta^{}_{23}=\pi/4$ and $\delta=3\pi/2$ at $\Lambda^{}_{\mu\tau}
= 10^{14} ~{\rm GeV}$ are fixed by the $\mu$-$\tau$ reflection symmetry.}
\end{figure}
%%%%%%%%%%%%%%%%%%%%%%%%%%%%%%%%%%%%%%%%%%%%%%%%%%%%%%%%%%%%%%%%%%%

\section{Summary}

While the nature of massive neutrinos (i.e., whether Dirac or Majorana)
remains an intriguing puzzle in particle physics,
it is largely believed that there should exist an
approximate $\mu$-$\tau$ reflection symmetry behind the observed
pattern of lepton flavor mixing. In this work we have studied such
a simple but interesting flavor symmetry for the Dirac neutrino
mass matrix, which can naturally
predict $\theta^{}_{23} = \pi/4$ and $\delta = \pi/2$ or
$3\pi/2$ in the standard parametrization of the PMNS matrix $U$.
Assuming the $\mu$-$\tau$ reflection symmetry is realized at a
superhigh energy scale $\Lambda^{}_{\mu\tau}$, we have investigated
how it is spontaneously broken via the one-loop RGEs
running from $\Lambda^{}_{\mu\tau}$ down to the Fermi scale
$\Lambda^{}_{\rm F}$ in two interesting scenarios: the MSSM and
the type-II 2HDM. Such quantum corrections to the neutrino
masses and flavor mixing parameters have been derived in a
perturbation approach, and an analytical
link has also been established between the Jarlskog invariants of leptonic
CP violation at $\Lambda^{}_{\mu\tau}$ and $\Lambda^{}_{\rm F}$.
In addition, we have illustrated the running behaviors of relevant
physical quantities by taking a few typical numerical examples in
the MSSM and the type-II 2HDM.

A particularly striking point of view associated with
this kind of study is that the octant of
$\theta^{}_{23}$, the quadrant of $\delta$ and the neutrino mass ordering
might be correlated with one another thanks to the RGE-triggered
breaking of $\mu$-$\tau$ reflection symmetry. We have illustrated this
observation both analytically and numerically by considering the massive Dirac
neutrinos in the MSSM and the type-II 2HDM, and found that these
two scenarios lead us to the opposite deviations of
$\theta^{}_{12}$, $\theta^{}_{13}$, $\theta^{}_{23}$ and $\delta$
from their corresponding values in the $\mu$-$\tau$ reflection
symmetry limit. Therefore, the future experimental data on the neutrino 
mass ordering and flavor mixing angles will allow us to make a 
choice between the MSSM and the type-II 2HDM, at least in this 
connection. Our results are also expected to be
useful for building explicit Dirac neutrino mass models and
explaining upcoming neutrino oscillation data.

\vspace{0.5cm}

We would like to thank S. Antusch, G.Y. Huang, S. Luo, J. Zhang, Z.H. Zhao 
and S. Zhou for useful discussions. This research work was supported
in part by the National Natural Science Foundation of China under
grant No. 11375207 and grant No. 11775231.

\end{document}